# Strong Coupling between Propagating Magnons and High-order gyrotropic Skyrmion modes

Zhihua Liu,[#] Kaile Xie,[#] and Fusheng Ma[*]

*Jiangsu Key Laboratory of Opto-Electronic Technology, Center for Quantum Transport and Thermal Energy Science, School of Physics and Technology, Nanjing Normal University, Nanjing 210046, China*

[#]These authors contributed equally to this work
[*]Correspondence to Fusheng Ma: phymafs@njnu.edu.cn

**Abstract**

Inspired by the recent achievements of the strong coupling between photon and spin excitation in magnetic systems, the coupling between the collective spin excitations of universal system and the local resonances of spin textures through direct exchange interaction is expected but not realized yet. In this work, we demonstrated the coherent coupling between propagating magnons and local skyrmion resonances. Besides the Rabbi coupling gap (RCG) in the frequency-field dispersion, a magnonic analog of polariton gap, *polaragnonic band gap* (PBG), is also observed in the frequency-wavenumber dispersion. The realization of coupling requires the gyrotropic skyrmion modes excited by propagating magnons to satisfy not only their quantum number should be larger than one but also their handness/chirality should be opposite to that of magnon. The observed PBG and RCG can be controlled to exist in different Brillouin zones (BZs) as well as BZ boundaries. The coupling strength can approach the strong regime by selecting the wavenumber of propagating magnon. Our findings could provide a pure magnonic platform for investigating coupling phenomena in quantum information and quantum technology.

# Introduction

Hybrid quantum systems that combine physical system with different advantages have been widely used in quantum computing, quantum sensing, and quantum communication [1–3]. Recently, the strong coupling between photon and the excitations in magnetic systems via dipolar interaction has been predicted and demonstrated. For instance, microwave cavity photon can couple with spin ensemble [4,5], magnon [6–13], spin textures (like, domain wall [14], vortex [15,16], and skyrmion [17–19]). Since magnons and spin textures can simultaneously excited in magnetic systems, one expects the coupling between collective spin excitations of universal system and local eigen-resonances of spin textures through direct exchange interaction, which may result in a polariton-like hybridization. To the best of our knowledge, such polaritonic spin texture system have remained unexplored.

Magnons have attracted widespread attention due to their advantages as information carriers in computing devices, such as no ohmic dissipation, multiple modulation methods, and so forth. In recent years, the research on the controllable excitation, transmission and modulation of magnon in the field of information processing and transport has gradually formed a new science field: magnonics [20–23]. Magnons with a right-handed chirality of various wavenumbers can propagate in uniformly magnetized waveguide as schematically shown in Fig. 1(b), and Fig. 1(e) is the corresponding frequency-wavenumber dispersion. The dispersion of magnons propagating in periodic magnetic structures, magnonic crystals (MCs), characterizes by magnonic bands (MBs) and magnonic band gaps (MBGs) owing to the interference of counterpropagating magnons as shown in Fig. 1(f). The periodicity of MCs could be introduced either by the spatial variation of structures [24–28] or the presence of spin textures [29–33].

Magnetic skyrmion is a topologically stable spin texture which has been experimentally observed in bulk crystals [34–39] and thin films [40–44]. Benefit from its topological stability, nanoscale size, and low driving current density, magnetic skyrmion shows high promising potential for ultra-high density storage and logic operations in functional spintronic devices [45–48]. As a quasi-particle, the collection excitations of skyrmion are comprised of out-of-plane breathing mode and in-plane gyrotropic modes [49–52]. The gyrotropic modes are indicated by a quantum number $l$, whose magnitude is integer representing the number of azimuthal periods, see Figs.1(c) and (d). The positive/negative sign of $l$ represents the counter-clockwise (CCW)/clockwise (CW) rotation, respectively [53–55].

The interactions between propagating magnon and spin textures have been reported as either magnons scattered by spin textures [54,56–59] or spin textures moved by magnons [60,61]. Actually, as a quasi-particle, spin texture can act as a resonator besides as a scatter. From the scatter point of view, the propagating magnons will be scattered by the magnetic skyrmions in a magnonic waveguide as shown in Fig. 1(a). The interference between the forward propagating magnon and the backward scattering magnon forms a standing wave resulting in the presence of Bragg type MBGs at Brillouin Zone (BZ) boundaries $k = n\pi/a$ [30–32,62–67]. From the resonator point of view, the $l^{th}$–order gyrotropic skyrmion modes can selectively excited depending on the wavelength of propagating magnon and the size of skyrmion. An open question is whether these locally gyrotropic modes can coherently couple to the propagating magnons with the presence of a polariton type band gap, which is similar to that in sonic crystal [68] and photonic crystal [69].

As a magnonic analogy of polariton [70–75], we denominate the quasi-particles formed as a result of the magnon-matter coupling as "*polaragnon*" and the resulting polariton band gap as *polaragnonic* band gap (PBG), as shown in Fig. 1(g). In this work, we proposed and numerically demonstrated the presence of PBGs formed through coupling between propagating magnons and high-order gyrotropic skyrmion modes. In contrast to Bragg type MBGs, the PBGs can appear not only at the BZ boundaries but also within BZs. It is found that the realization of magnon-skyrmion coupling is codetermined by the quantum number and chirality of the gyrotropic skyrmion modes. Furthermore, the strong-coupling regime could be approached by manipulating the wavenumber of propagating magnons for different high-order gyrotropic skyrmion modes. The realization of magnon-skyrmion coupling could provide an all-spin magnonic platform for investigating quantum information and technology in quantum optics.

**Results**

Fig. 2(a) is the calculated frequency-wavenumber ($f$-$k_x$) dispersion in the first two BZs for magnon propagating along $x$ direction in the magnonic waveguide with dimension of 6400 nm × 40 nm × 0.4 nm at $H_z$ = 185.3 mT. Two characteristic frequency band gaps are observed at $k_x = \pi/a$ and $1.5\pi/a$, respectively. The band gap at $k_x = \pi/a$ is the "zone folding" type, which is attributed to the Bragg scattering of magnon by the periodically arranged skyrmions. While, the band gap at $k_x = 1.5\pi/a$ is the anticrossing type, as shown in Fig. 2(b). Since the observed anticrossing gap appears at the crossing point of bare magnon mode (blue dot-dash line) and locally gyrotropic skyrmion mode (orange dot-dash line), we attribute this gap to the repulsive coupling between magnon and skyrmion. Similar gap has been observed in phononics [68] and photonics [69] as polariton gap, as an analog, we call the gap at $k_x = 1.5\pi/a$ as polaragnonic band gap (PBG). Additionally, a Rabi coupling gap (RCG) in the frequency-field ($f$-$H_z$) dispersion can also characterize the coupling between different elementary excitations [76]. We also calculated the $f$-$H_z$ dispersion for $k_x = 1.5\pi/a$ as shown in Fig. 2(c), in which the hallmarks of coupling, anticrossing and linewidth evolution, is clearly observed.

To find out the quantum number $l$ of the gyrotropic skyrmion mode participating in the coupling with magnon, we calculated six typical profiles of the skyrmion modes at selected fields and frequencies as indicated by ① - ⑥ in Fig. 2(c). For modes ① - ② as shown in the first column of Fig. 2(d), they are the bare propagating magnon mode which can excite the $|l = -1\rangle$ gyrotropic skyrmion mode. The $|l = -1\rangle$ mode is only localized around the skyrmion core and its coupling with magnon is negligible. For the convenience of description, the propagating magnon mode will be hereafter represented by the $|l = -1\rangle$ mode. For the modes ③ - ④ as shown in the second column of Fig. 2(d), they are the high-order $|l = 2\rangle$ gyrotropic skyrmion mode with a clockwise rotation sense, which is opposite to the chirality of magnon. (A dark mode [19] that is not easily excited alone). For modes ⑤ - ⑥ as shown in the third column of Fig. 2(d), they are the upper/lower branch of the polaragnon which is the quasi-particle of the coupling between the $|l = -1\rangle$ and $|l = 2\rangle$ modes. The hybridization of various modes $|l\rangle$ can be represented as $|\psi_{l_\mu, l_\nu}\rangle = \mu |l = l_\mu\rangle + \nu |l = l_\nu\rangle$ where $l_\mu$ and $l_\nu$ is the quantum number of the two uncoupled modes, and the positive $\mu$ and $\nu$ are the weight coefficients. Then, the coupling between the $|l = -1\rangle$ and $|l = 2\rangle$ modes can be described as $|\psi_{-1, 2}\rangle$. For the upper polaragnon branch (PB), the mode ① is dominated by the $|l = -1\rangle$ mode. The mode ⑤ is the hybridization of the $|l = -1\rangle$ and $|l = 2\rangle$ modes with equal contributions. The mode ④ is dominated by the $|l = 2\rangle$ mode. For the lower PB, the mode ③ is dominated by the $|l = 2\rangle$ mode. The mode ⑥ is the hybridization of the $|l = 2\rangle$ and $|l = -1\rangle$ modes with equal contributions. The mode ② is dominated by the $|l = -1\rangle$ mode. A phenomenological explanation of the hybridization of the $|l\rangle$ modes can be understood from spatial distribution of $\Delta m_z$ along the circumference of skyrmion for the $|l = -1\rangle$, $|l = 2\rangle$, and $|\psi_{-1, 2}\rangle$ modes. For instance, Fig. 2(e) shows the $\Delta m_z$ of modes ①, ③, and ⑤ along the dashed circles in Fig. 2(d). The mode ⑤ can be obtained from the superposition of modes ① and ③ by considering their wavelength, amplitude, and phase as schematically revealed in Fig. 2(f).

Next, we studied the temporal variation of the magnetization components $\Delta m_x$ and $\Delta m_z$

for various frequencies $f_1 - f_5$ as marked in Fig. 2(a). For clarity, a part of the magnonic waveguide, only the first five skyrmions from the magnon generation area, is displayed in Fig. 3. The static magnetization distributions $m_x$ (left) and $m_z$ (right) at $H_z = 185.3$ mT are shown in Fig. 3(a). And the snapshots of dynamic magnetization $\Delta m_x$ (left) reflecting the propagation properties of magnons and $\Delta m_z$ (right) reflecting the quasi-particle properties of skyrmions for different frequencies are shown in Figs. 3(b)-(c). Magnon of $f_1 = 40.3$ GHz located in the first MB can propagate along the waveguide and excite the $|l = -1\rangle$ gyrotropic skyrmion mode as shown in Fig. 3(b). It is worth to point out that the gyrotropic motion of neighboring skyrmions exhibits a fixed phase difference which indicates the magnon-mediated correlation between skyrmions [77]. In contrast, magnon of $f_2 = 45.7$ GHz located in the first MBG cannot propagate and dissipate quickly due to the scattering of magnons by skyrmions [30–32,62–67].

Now, we consider magnons of the frequencies located in the PB and PBG from the $|\psi_{-1, 2}\rangle$ coupling. For magnon of $f_3 = 50.6$ GHz located in the lower PB, it can propagate along the waveguide as shown in the left column of Fig. 3(c). The propagating magnons can also excite the skyrmions into gyrotropic motion. Interestingly, the skyrmion motion is neither $|l = -1\rangle$ nor $|l = 2\rangle$ but a hybridized $|\psi_{-1, 2}\rangle$ mode as shown in the right column of Fig. 3(c), see also movie 1 in Supplemental Material [78]. A similar phenomenon is also observed for magnons of frequency $f_5 = 54.9$ GHz located in the upper PB as shown in the bottom panel of the Fig. 3(c). For magnons of $f_4 = 52.6$ GHz located in the PBG, they cannot propagate as shown in the left column of Fig. 3(c). Here, the magnons in the PBG dissipate slower than that in the MBG, see the movie 2 in Supplemental Material [78], which is similar to the surface phonon propagation in locally resonant band gap [79]. While the magnons can excite the $|l = 2\rangle$ grotropic skyrmion mode before they annihilate. It should be pointed out that the gyotropic motion of skyrmion is in the clockwise manner opposite to the chirality of magnon.

So far, we have understood that the coupling between the propagating magnons and high-order gyrotropic skyrmions can result in the presence of PBG in the $f$-$k_x$ dispersion and RCG in the $f$-$H_z$ dispersion as shown in Figs. 2(b) and (c) for $|\psi_{-1,\,2}\rangle$ polaragnon. Since the order of the excited gyrotropic skyrmion mode is highly sensitive to the wavelength of magnon and the skyrmion size, the magnon-skyrmion coupling is expected to be related to $k_x$ and $H_z$. Thus, we calculated the dispersions of polaragnon under three selected fields $H_z$ = 174.6, 195.3, and 200 mT as shown in Fig. 4(a). It is evident that with the increasing of field $H_z$, the PBG appears at higher frequency and larger wavenumber. And for three typical wavenumbers $k_x = \pi/a$, $1.75\pi/a$, and $2\pi/a$, the calculated $f$-$H_z$ dispersions for the $|\psi_{-1,\,2}\rangle$ polaragnon is shown in Fig. 4(b). It is observed that the RCG changes obviously with $k_x$. To evaluate the Rabi splitting, a coupling strength $g$ is defined as a half of the frequency difference between the two hybridization modes when they have the similar intensity, indicated by the spectrum in Fig. 4(b). For the $|\psi_{-1,\,2}\rangle$ polaragnon, the dependence of PBG and $g$ on $H_z$ and $k_x$ are summarized in Fig. 4(c). The width of PBG firstly increases from 0.38 to 0.799 GHz with increasing $H_z$ from 174.6 to 195.3 mT, and then decreases to zero with increasing $H_z$ to 203 mT. Similarly, the $g$ firstly increases from 0.5 GHz to 0.9 GHz with increasing $k_x$ from $\pi/a$ to $1.75\pi/a$, and then decreases to zero with increasing $k_x$ to $2\pi/a$. It is worthy to pointing out that strong coupling regime, coupling strength exceeding dissipation rates, could be approached as labeled by the stars in Fig. 4(c).

## Discussion and Conclusion:

We have discussed the coupling between propagating magnon (the $|l=-1\rangle$ skyrmion mode) and the high-order $|l=2\rangle$ skyrmion mode. Actually, such kind of magnon-skyrmion coupling is also observed for even higher order skyrmion modes, like $|l=3\rangle$, $|l=4\rangle$, and so forth. For instance, the description of the $|\psi_{-1,\,3}\rangle$ coupling is shown as Figs. S1-S3 in Supplemental Material [78]. In contrast to the Bragg band gaps, the PBG can exist in the BZs as well as the BZ boundaries. For the $|\psi_{-1,\,2}\rangle$ coupling, the PBG appears in the 2$^{nd}$ BZ.

By adjusting the period of skyrmion, the PBG can also be observed in other BZs as shown in Fig. S4 in Supplemental Material [78]. Therefore, the location of PBG can be manipulated by magnetic field in the same BZ and by skyrmion period in different BZs. In addition, the resulting magnon-skyrmion coupling could be tuned in strength and position by engineering multiple parameters, such as *DMI*, $K_u$, and $M_s$.

In conclusion, we introduced the coupling between the propagating magnon mode and high-order gyrotropic skyrmion mode with left-handed chirality. The coupling is accompanied by the appearance of PBG which can exist within different BZs as well as BZ boundaries, and its position can be tuned by the magnetic field, skyrmion period and other material parameters. Furthermore, the coupling strength represented by RCG in *f-$H_z$* dispersion is highly dependent on the wavenumber of magnon. For the selected wavenumber of propagating magnon, the coupling strength can approach the strong regime.

## Acknowledgements

This work was supported by the National Natural Science Foundation of China (Grant Nos. 12074189 and 11704191).

# Figures

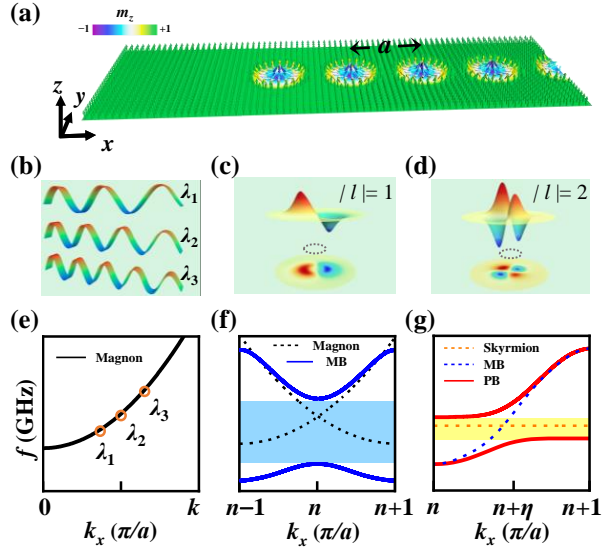

**FIG. 1.** (a) Schematic of magnonic waveguide with uniform magnetization (left) and periodic array of skyrmions with lattice constant $a = 40$ nm (right). Propagating magnons are excited from the left end of the waveguide. (b) Snapshot of magnons propagating along uniformly magnetized waveguide with various wavelength $\lambda$. Profiles of gyrotropic skyrmion modes of different quantum numbers (c) $|l| = 1$ and (d) $|l| = 2$, represented by the temporal variation of the magnetization component $\Delta m_z$. Dashed circles indicate the perimeter of skyrmions. Frequency-wavenumber dispersion relations of magnons propagation in magnonic waveguide with uniform magnetization (e) and periodic array of skyrmions (f) and (g). (f) Bragg MBG only appears at the BZ boundaries as indicated by blue shaded area. Solid lines denote MBs, dot-dash lines represent the bare magnons mode. (g) PBG appears at the crossing point of bare magnonic band (blue dot-dash line) and bare high-order gyrotropic skyrmion modes (orange dot-dash line) as indicated by yellow shaded area. Solid lines denote upper/lower PBs. $\eta$ satisfies $0 \leq \eta \leq 1$ indicating that PBG can occur in BZs as well as BZ boundaries.

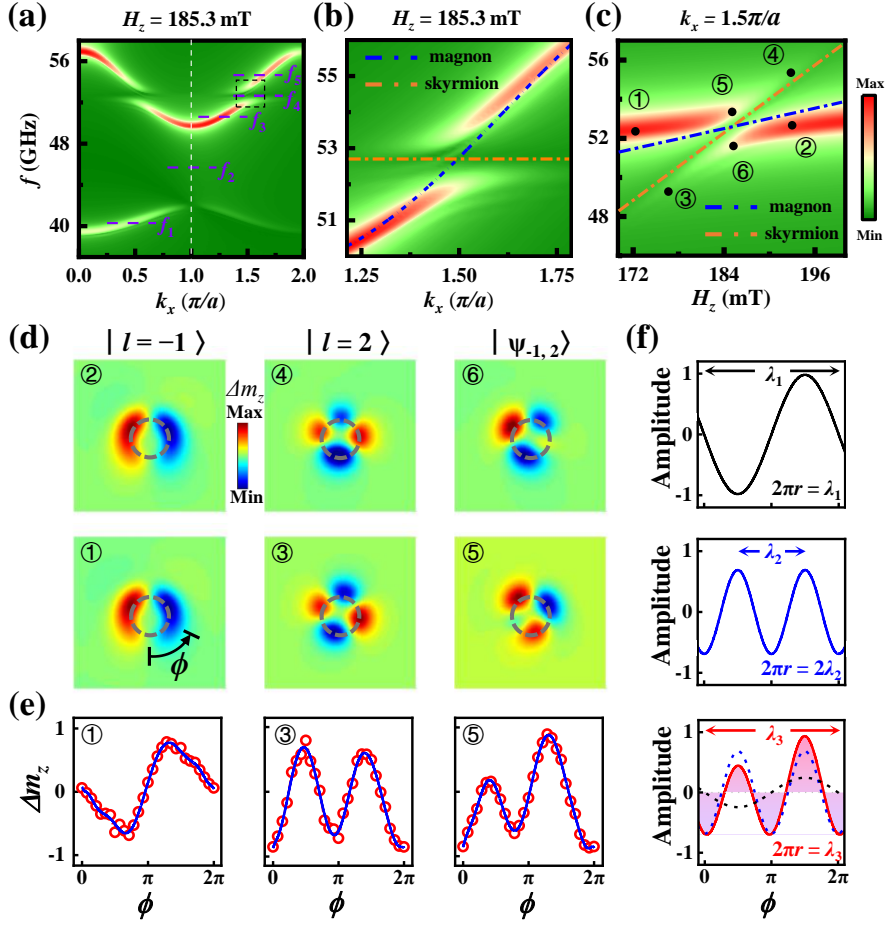

**FIG. 2.** Color plots of dispersion relation of magnons in magnonic waveguide with periodic array of skyrmion of $a = 40$ nm: (a) $f$-$k_x$ dispersion for $H_z = 185.3$ mT, (b) local $f$-$k_x$ dispersion as indicated by square in (a), and (c) $f$-$H_z$ dispersion for $k_x = 1.5\pi/a$. The blue and orange dot-dash lines represent the bare magnons mode and bare $|l = 2\rangle$ gyrotropic skyrmion mode, respectively. (d) Profiles of gyrotropic skyrmion modes of various $f$ and $H_z$ as labeled numbers in (c). Dashed circles indicate the perimeter of skyrmions. (e) Normalized $\Delta m_z$ as a function of azimuthal angle $\phi$ along the perimeter of skyrmion extracted from modes ①, ③, ⑤ in (d). (f) Schematically explanation of wave $\lambda_3$ superimposed by waves $\lambda_1$ and $\lambda_2$ by considering their relative wavelength, amplitude and phase.

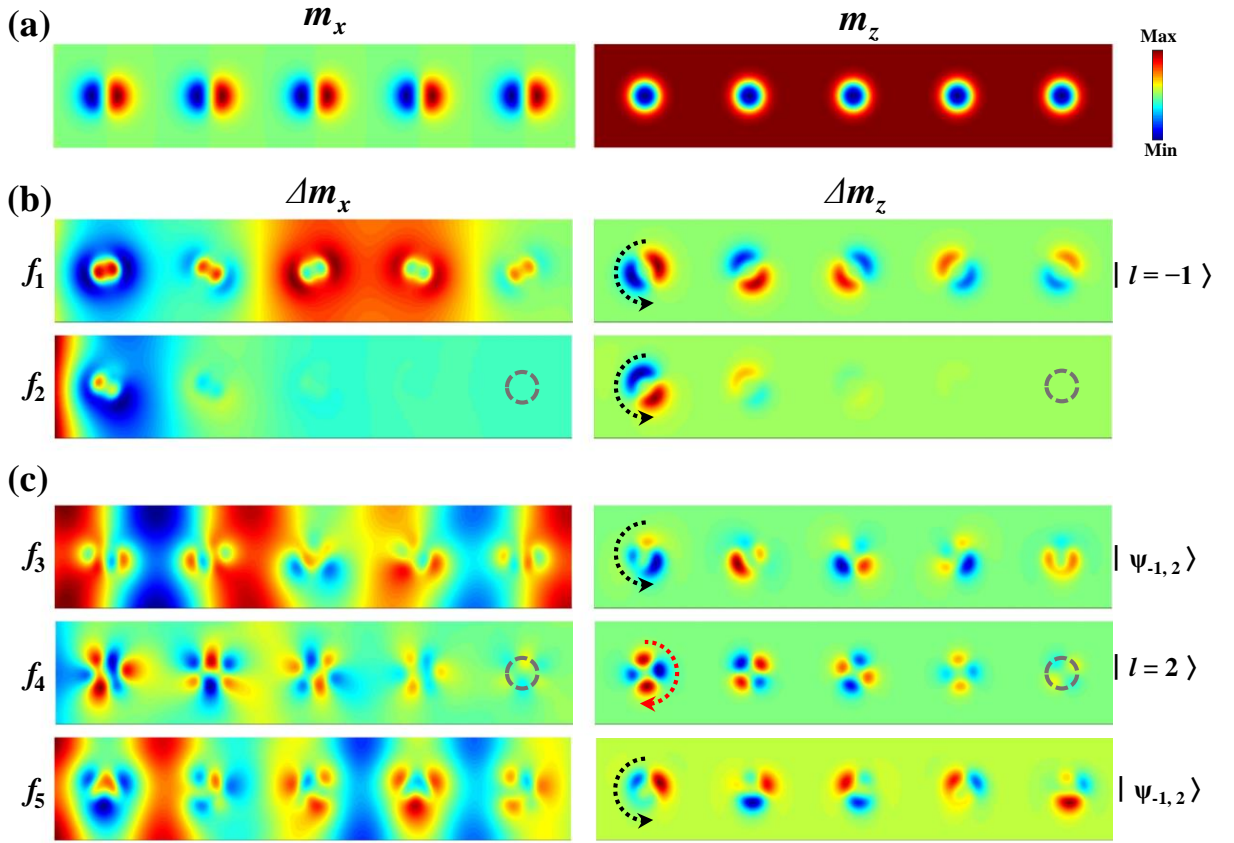

**FIG. 3.** (a) Spatial distribution of $m_x$ (left) and $m_z$ (right) at $H_z = 185.3$ mT. Colors correspond to the magnetization component $m_x$ (left) and $m_z$ (right). (b)-(d) Snapshots of temporal variation of the magnetization component $\Delta m_x$ (left column) and $\Delta m_z$ (right column) for various frequencies $f_1$-$f_5$ as labelled in Fig. 2(a). (b) $f_1$ located in the first MB and $f_2$ located in the first MBG. (c) $f_3$ located in the lower PB, $f_4$ located in the PBG, and $f_5$ located in the upper PB for the $|\psi_{-1,2}\rangle$ coupling. Dashed circles indicate the perimeter of skyrmions. Dashed arrows indicate the rotation direction.

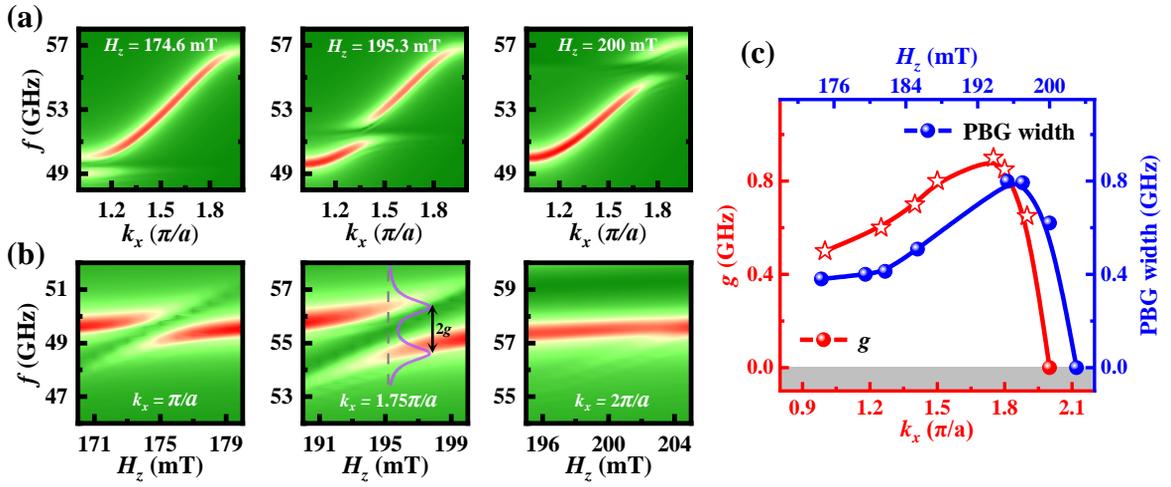

**FIG. 4.** Color plots of dispersion of $|\psi_{-1,2}\rangle$ coupling: (a) $f$-$k_x$ dispersion for $H_z$ = 174.6, 195.3, and 200 mT; (b) $f$-$H_z$ dispersion for $k_x$ = $\pi/a$, 1.75$\pi/a$, and 2$\pi/a$. (c) Coupling strength $g$ as a function of wavenumber $k_x$ (left axis) and PBG width as a function of magnetic field $H_z$ (right axis) for the $|\psi_{-1,2}\rangle$ coupling. Stars indicate magnons-skyrmion coupling in the strong regime.